\documentclass[preprint,superscriptaddress,preprintnumbers,showpacs]{revtex4}
\usepackage{amssymb}
\usepackage{graphicx}
\usepackage{dcolumn}
\usepackage{bm}

\begin{document}

\newcommand*{\PKU}{School of Physics and State Key Laboratory of Nuclear Physics and
Technology, Peking University, Beijing 100871}\affiliation{\PKU}
\newcommand*{\NTU}{Department of Physics and Center for Theoretical Sciences, National Taiwan University, Taipei 10617}\affiliation{\NTU}
\newcommand*{\IPPC}{Institute of Particle Physics and Cosmology, Department of Physics,
Shanghai JiaoTong University, Shanghai 200240}\affiliation{\IPPC}
\newcommand*{\CHEP}{Center for High-Energy
Physics, Peking University,
Beijing 100871}\affiliation{\CHEP}

\title{Quantization of Black Holes}

\author{Xiao-Gang He}\email{hexg@phys.ntu.edu.tw}\affiliation{\PKU}\affiliation{\NTU}\affiliation{\IPPC}\affiliation{\CHEP}
\author{Bo-Qiang Ma}\email{mabq@phy.pku.edu.cn}\affiliation{\PKU}\affiliation{\NTU}\affiliation{\CHEP}

\begin{abstract}
We show that black holes can be quantized in an intuitive and
elegant way with results in agreement with conventional knowledge of
black holes by using Bohr's idea of quantizing the motion of an
electron inside the atom in quantum mechanics. We find that
properties of black holes can be also derived from an Ansatz of
quantized entropy $\Delta S=4\pi k {\Delta R / \lambdabar}$, which
was suggested in a previous work to unify the black hole entropy
formula and Verlinde's conjecture to explain gravity as an entropic
force. Such an Ansatz also explains gravity as an entropic force
from quantum effect. This suggests a way to unify gravity with
quantum theory. Several interesting and surprising results of black
holes are given from which we predict the existence of primordial
black holes ranging from Planck scale both in size and energy to big
ones in size but with low energy behaviors.
\end{abstract}



\maketitle


The quantization of black holes is one of the important issues in
physics~\cite{'tHooft:1984re}, and there has been no
satisfactory solution yet. Black holes are characterized by the
no-hair theorem, which states that once a black hole achieves a
stable condition regardless how it is formed, it has only three
independent physical properties: mass $M$, charge $Q$, and angular
momentum $J$. This theorem is also called three-hair theorem
sometimes.

Once the mass $M$ of a black hole with no charge and zero angular
momentum is given, its other quantities should be completely fixed,
as there is only one independent parameter from the no-hair theorem.
Therefore the Schwarzschild radius $R=2 G M/ c^2$, the surface area
$A=4\pi R^2$, and the Compton wavelength $\lambdabar= \hbar / M c$
of a black hole cannot be considered as independent parameters from
each other but only one quantity in essence. One elegant equality
among possible relations is
\begin{equation}
R \lambdabar =2 l_{P}^2, \label{Rl}
\end{equation}
where $l_P=\sqrt{G\hbar /c^3}=1.61624(8)\times 10^{-35}\,$m is the
Planck length, $\hbar$ is the Planck constant, and $c$ is the speed
of light, and $G$ is the universal gravitational constant called
Newton constant. The relation ($\ref{Rl}$) is satisfied for any mass
$M$, so we take it as a given fact without further inquiry about its
rationality.

As there is only one parameter for a black hole with no charge and
angular momentum, it is natural to take the black hole as a sphere
with Schwarzschild radius $R$ as its boundary. Since nothing inside the
sphere can be known for an outside observer, we should consider all
of the information of the black hole as recorded on the surface,
which is called event horizon or holographic screen from the
holographic principle~\cite{thooft,susskind}. There have been many
studies of the physical properties of black holes.
Bekenstein~\cite{Bekenstein} conjectured that a black hole entropy
is proportional to the area $A$ of its event horizon divided by the
Planck area $l_{P}^2$, and later the entropy formula of a black hole
was found to be~\cite{Bardeen,Hawking}
\begin{equation}
S=\frac{k A}{4 l_{P}^2}, \label{entropy}
\end{equation}
where $k$ is the Boltzmann constant. The black holes create and emit
particles as if they were black bodies with temperature
\begin{equation}
T=\frac{\hbar a}{2\pi c k}, \label{Temp}
\end{equation}
where $a$ denotes the surface gravity of the black
hole~\cite{Hawking} or the acceleration of a test particle at the
horizon~\cite{Unruh}.

Due to the particle-wave duality property in quantum theory, any
particle with energy $E_m=m c^2$ should also behave like a wave with
the de~Broglie wavelength $\lambda_m^d=h /m v$, which can be also
expressed as $\lambdabar_m^d=\hbar /m v$, where $\hbar=h/2\pi$ and
$v$ is the velocity of the particle.
In Bohr's theory of the one-electron atom, the condition for the
electron-wave to be stable around its circular orbit with radius $r$
is given by
\begin{equation}
2\pi r=n \lambda_m^d,  ~~~~~\text{or} ~~~~~~~r=n \lambdabar_m^d,
\end{equation}
which corresponds to the condition for the electron to be in a
standing wave state with $n$ integer nodes. The electron of the atom
is thus quantized in discrete energy states. The way to change the
electron from one state to another is to emit or absorb a photon
with energy to be equal to the energy difference between the two
states.

Similarly, we can naturally speculate that a black hole with mass
$M$ also behaves like a wave with the Compton wavelength $\lambdabar
=\hbar /M c$. From a classical viewpoint, nothing, including the
wave, can escape from the black hole horizon, therefore the wave of
the black body should also propagate within the sphere on the
surface. One may first speculate that the condition for the black
hole to be stable is also that its wave to be in a standing wave
state, i.e.,
\begin{equation}
2\pi R=\tilde{n} \lambda,  ~~~~~\text{or} ~~~~~~~R=\tilde{n}
\lambdabar.
\end{equation}
We should notice that the wave is not moving on a circular orbit,
but on a sphere surface, so the number $\tilde{n}$ of nodes is
adopted as an even number, hence we take $\tilde{n}=2n$ and get
\begin{equation}
\pi R=n \lambda,  ~~~~~\text{or} ~~~~~~~ R= 2 n \lambdabar,
\label{BHquan}
\end{equation}
where $n$ is an integer number now, no matter it is even or odd.

By applying the equality (\ref{Rl}) in the above condition to
eliminate $\lambdabar$, we get the Schwarzschild radius of the black
hole
\begin{equation}
R_n=2\sqrt{n}l_P,
\end{equation}
which means that black holes are quantized in discrete states with
radius $R_n$. Thus we easily obtain the other quantities of black
holes, i.e., the energy
\begin{equation}
E_n=M_n c^2={R_n c^4 \over 2 G}=\sqrt{n}M_Pc^2, \label{energy}
\end{equation}
where the Planck mass $M_P=\sqrt{\hbar c/G}=1.22089(6)\times
10^{19}\,\text{GeV}/\text{c}^2=2.17644(11)\times
10^{-8}\,\text{kg}$; the surface area
\begin{equation}
A_n=4\pi R_n^2= 16\pi n l_P^2;
\end{equation}
and the Compton length
\begin{equation}
\lambdabar_n={\hbar \over M_n c}={l_P \over \sqrt{n}}.
\label{compton}
\end{equation}
The Planck scale relations, i.e., $M_P l_P=\hbar/c$ and
$M_P/l_P=c^2/G$, are useful for the derivation of the above
expressions.

When the black hole is quantized, the surface gravity $a$ is also
quantized by
\begin{equation}
a_n={G M_n\over R_n^2}={c^2\over 4 \sqrt{n} l_P }\;.
\end{equation}
This also leads to a quantized temperature $T$
\begin{equation}
T_n={\hbar a_n \over 2 \pi c k }={M_P c^2\over 8 \pi \sqrt{n} k}\;.
\end{equation}

From above we see that black holes have been quantized in a simple
and elegant way with interesting and surprising results,
acceleration and temperature are also quantized.

It is most interesting to notice that the entropy formula is now given by
\begin{equation}
S=4\pi k n,
\label{q-entropy}
\end{equation}
which is quantized as proposed in Ref.~\cite{HeMa} that the entropy
change $\Delta S$ for a 2 dimensional holographic screen is given by
$4\pi k$, for the purpose to reconcile the black hole energy formula
with the Verlinde conjecture of the entropic force
idea~\cite{Verlinde}.

Reversely, if we make the Ansatz that the entropy change of a black
hole with its horizon radius $R$ change $\Delta R$ is, according to
the rule suggested in Ref.~\cite{HeMa},
\begin{equation}
\Delta S=2\pi k D {\Delta l \over \lambdabar}, \label{ansatz}
\end{equation}
where $\Delta l$ is a linear displacement to cause
entropy change and $D$ is the dimensional degree of freedom of the
objects under consideration, and apply to a black hole, we obtain
\begin{equation}
\Delta S=4\pi k {\Delta R \over \lambdabar}=4\pi k { R \Delta R
\over 2 l_P^2}={k \Delta A \over 4 l_P^2}.
\end{equation}
Up to an integration constant
\begin{equation}
S=4\pi k { R^2 \over 4 l_P^2} + S_0={k A \over 4
l_P^2} + S_0.
\end{equation}
The constant can be chosen to be zero in consistent with intuitive
expectation that no area no entropy for black hole.
We then obtained the black hole entropy formula (\ref{entropy}),
consistent with that obtained from classical
considerations~\cite{Bekenstein,Bardeen,Hawking}.

If the entropy change on the black hole holographic screen is due to
a test particle of Compton wavelength $\lambdabar_m$ moving $\Delta
x$ towards the black hole, according to the Verlinde conjecture of
the entropic force idea~\cite{Verlinde}, one has $\Delta S = 2 \pi k
\Delta x/\lambdabar_m$. If $\Delta x/2\lambdabar_m = \Delta
R/\lambdabar$ is quantized as suggested in Ref.\cite{HeMa}, one can
then obtain (\ref{q-entropy}). From there, one can go backwards in
our previous derivations to obtain the same quantization rules for
the radius, the black hole wavelength, the mass spectrum and the
temperature.


This tells us that one can relate black hole quantization with quantization of entropic
change~\cite{HeMa}, which can also explain gravity as an entropic
force~\cite{Verlinde,Padma,Padma2}. The quantization of black holes
found in this work, however,  should not be interpreted as only a
support of gravity as an entropic force, it also suggests a new way
to unify gravity with quantum theory.

Now we discuss some of the physical implications and predictions can
be obtained from the above results. First, the most small stable
black hole is obtained for $n=1$. The Schwarzschild radius $R_1=2
l_P$ doubles that of the Planck length $l_P$, the energy $E_1$ is
just the Planck mass $M_P c^2$, and the Compton length just equals
to the Planck length $\lambdabar_1=l_P$. Thus the most small black
hole is of the Planck scale both in size and energy as
expected~\cite{Bekenstein:1974jk}. It also supports the proposal for
the existence of primordial black holes. These black holes range
from mini black holes of Planck scale to very big ones with large
$n$. Their distribution spectrum can be calculated and consequences
can be also predicted from the results in this work.

From the quantized energy formula (\ref{energy}) we get the energy
difference between two nearby states
\begin{equation}
\Delta E=E_{n+1}-E_n=(\sqrt{n+1}-\sqrt{n})M_P c^2={M_P c^2 \over
\sqrt{n+1}+\sqrt{n}}\;. \label{Denergy}
\end{equation}
Note that the largest difference is $(\sqrt{2}-1)M_P c^2$ and
$\Delta E$ decreases with the increase of $n$. When $n$ is very
large, one gets
\begin{equation}
\Delta E\approx {M_P c^2\over 2 \sqrt{n}}\to 0 ~~~~ \text{when}
~~~~n \to \infty\;.
\end{equation}
Therefore the energy change between two nearby states for a big
black hole approaches to zero, i.e., it possesses also low energy
continuum behavior. In this case, the black hole temperature also approaches to zero.
This means that a black hole can absorb a
particle with small energy by transition between nearby states or a
big one by transition between far-away states.

It is natural to ask what would happen during the transition between
two states. Similar to the Bohr's theory of one-electron atom, we
can also predict the quantum emission and absorbtion of a charge
neutral particle ranging from Planck scale to low scale for the
specific case discussed in this work. As the situation considered in
this work is a black hole with no angular momentum, one may still
not be possible to identify the particle as photon yet. It is
possible to make extension of this study to general situation with
both charge and angular momentum being considered. In that case the
black hole can emit or absorb particles or objects with any mass and
spin or total angular momentum. However, the mini black holes can
only emit or absorb particles with finite energy and definite
quantum numbers, in similar to the case of atomic spectrum in
quantum mechanism.

Our approach shows that the quantization of black holes of
gravitational in nature can be performed in the same way as that for
the electron motion inside the atom from quantum theory. It leads to
interesting results on black holes as shown in this work. The
entropic framework~\cite{HeMa,Verlinde} indeed offers new
perspectives on quantum properties of gravity beyond classical
physics. The quantization of black holes found in this work,
however,  should not be interpreted as only a support of gravity as
an entropic force, it also suggests a new way to unify gravity with
quantum theory. We anticipate that these ideas will lead to new
understanding and perspectives on gravity.
\\

\begin{acknowledgments}
This work is partially supported by NSC, NCTS, NSFC (Nos.~10721063,
10975003). BQM acknowledges the support of the LHC physics focus
group of NCTS and warm hospitality from W.-Y. Pauchy Hwang during
his visit of NTU. He also acknowledges discussions with Jiunn-Wei
Chen, Pisin Chen, Tzihong Chiueh, George Wei-Shu Hou, Shin Nan Yang,
and Hoi-Lai Yu.
\end{acknowledgments}

\end{document}